\documentclass[12pt]{article}
 \usepackage{epsfig}
  \usepackage{wrapfig}
   \usepackage{graphicx}
 \usepackage{color}

\begin{document}

\begin{titlepage}
\begin{center}

{\Large\bf New analysis concerning the strange quark\\[2mm]
polarization puzzle}

\end{center}
\vskip 2cm
\begin{center}
{\bf Elliot Leader}\\
{\it Imperial College London\\ Prince Consort Road, London SW7
2BW, England }
\vskip 0.5cm
{\bf Alexander V. Sidorov}\\
{\it Bogoliubov Theoretical Laboratory\\
Joint Institute for Nuclear Research, 141980 Dubna, Russia }
\vskip 0.5cm
{\bf Dimiter B. Stamenov \\
{\it Institute for Nuclear Research and Nuclear Energy\\
Bulgarian Academy of Sciences\\
Blvd. Tsarigradsko Chaussee 72, Sofia 1784, Bulgaria }}
\end{center}

\vskip 0.3cm
\begin{abstract}
\hskip -5mm The fact that analyses of \emph{semi-inclusive} deep
inelastic scattering suggest that the polarized strange quark
density $\Delta s(x) + \Delta \bar{s}(x)$ is positive in the
measured region of Bjorken $x$, whereas all analyses of
\emph{inclusive} deep inelastic scattering yield significantly
negative values of this quantity, is known as the ``strange quark
polarization puzzle". We have analyzed the world data on inclusive
deep inelastic scattering, including the COMPASS 2010 proton data
on the spin asymmetries, and for the first time, the new extremely
precise JLab CLAS data on the proton and deuteron spin structure
functions. Despite allowing, in our parametrization, for a
possible sign change, our results confirm that the inclusive data
yield significantly negative values for the polarized strange
quark density.


\end{abstract}

\end{titlepage}

\newpage
\setcounter{page}{1}

\section{Introduction}

In the absence of neutrino reactions on a polarized target, the
inclusive polarized deep inelastic lepton-hadron reactions
determine only the sum of quark and anti-quark polarized parton
density functions (PDFs), $\Delta q(x) + \Delta \bar{q}(x)$, and
provide no information at all about the individual polarized quark
and anti-quark densities. All analyses of the polarized
\emph{inclusive} (DIS) data have produced results for the
polarized strange quark density function, $\Delta s(x) + \Delta
\bar{s}(x)$, which are significantly \emph{negative} for all
values of $x$ (see for instance \cite {{LSS07},{groups}}), in
contradiction to the positive values obtained from combined
analyses of inclusive and semi-inclusive deep inelastic scattering
data \cite{{DSSV},{LSS10}} using de Florian, Sassot, Stratmann
(DSS) fragmentation functions (FFs) \cite{DSS}. This problem is
known as the strange quark polarization puzzle. It was shown
\cite{LSS11}, however, that the polarized strange quark density is
very sensitive to the kaon fragmentation functions, and if the set
of Hirai, Kumano, Nagai, Sudoh (HKNS) fragmentation functions
\cite{HKNS} is used, the polarized strange quark density obtained
from the combined analysis turns out to be negative and well
consistent with values obtained from the pure deep inelastic
scattering analyses. Since it has turned out that neither the HKNS
nor the DSS FFs are consistent with the recent HERMES data on pion
and kaon \emph{multiplicities} \cite{HERMES_mult}, one can
conclude now that the values for the polarized strange quark
density $\Delta s(x) + \Delta \bar{s}(x)$ determined from the
combined analyses \cite{{DSSV},{LSS10}} and \cite{LSS11} of the
inclusive and semi-inclusive DIS data, cannot be correct. On the
other hand, a disadvantage of the QCD analyses of the pure
inclusive polarized DIS data is that in all of them simple input
parametrizations for the polarized strange quark density, which do
not permit a sign change of the density, have been used. Note that
the value of the first moment of the polarized strange quark
density must be negative.  This follows from the experimental
values for $\Delta\Sigma$, the spin carried by all the quarks, and
for $a_8=3F-D$, where $a_8$ is the 8th component of the axial
Cabibbo current, with  constants $F$ and $D$ determined from
hyperon $\beta$ decays. Thus if $\Delta s(x) + \Delta \bar{s}(x)$
is positive for medium values of $x$, it has to be negative at
small values of $x$, implying that there should be a sign change.
The previous simple input parametrizations were used because the
data did not allow a reasonable determination of the parameters
responsible for the sign change \cite{note}. The situation has now
changed.

In this paper we present a next-to-leading order (NLO) QCD
analysis of the polarized inclusive DIS data including in the
world data set the recent very precise JLAB CLAS data on the
proton and deuteron spin structure functions \cite{CLAS14}. The
aim of our analysis is to answer the question if it is possible,
in the presence of the new CLAS data, to determine the polarized
strange quark density $\Delta s(x,Q^2) + \Delta \bar{s}(x,Q^2)$
using a more general input parametrization which allows for a sign
change. Compared with our last fit to inclusive DIS data
\cite{LSS07}, the following changes are made:

~~~(i) We use now more general input parametrizations for the sum
of quark and anti-quark polarized PDFs $\Delta q(x) + \Delta
\bar{q}(x)$ instead of the valence and sea quark densities. In
particular, for the polarized strange quark density, allowance is
made for a sign change of the density.

~~~~(ii) We do not make any assumptions about the polarized light
sea quark densities $\Delta \bar{u}(x)$ and $\Delta \bar{d}(x)$
which have been used in all previous analyses, because as was
stressed above only the sums $\Delta q(x,Q^2) + \Delta
\bar{q}(x,Q^2)$ can be extracted from the data, and the
assumptions made cannot be tested. Note here that in contrast to
the light sea quark densities, the strange quark density $(\Delta
s + \Delta \bar{s})(x,Q^2)$ can be well determined from the
inclusive data if they are sufficiently precise.

In addition, the COMPASS proton data on the spin asymmetries
\cite{COMPASSp}, which were not available at the time of our last
analysis of the inclusive DIS data \cite{LSS07}, have  also been
included.

\section{Results of Analysis}

In this section we will present and discuss the results of our new
NLO QCD fit to the present world data on polarized inclusive DIS
adding to the old data set (\cite{EMC}-\cite{CLAS06}), used in our
previous analysis \cite{LSS07}, the COMPASS proton data
\cite{COMPASSp} and the new CLAS data \cite{CLAS14}. The data used
(902 experimental points) cover the following kinematic region:
$\{0.005 \leq x \leq 0.75,~~1< Q^2 \leq 62~GeV^2\}$. Note that for
the CLAS data a cut $W > 2~GeV$ was imposed in order to exclude
the resonance region.

The method used is the same as in our previous QCD analysis of the
inclusive DIS data (see \cite{LSS07} and the references therein).
The main difference, as was mentioned in the Introduction, is that
we use now input parametrizations at $Q_0^2=1~GeV^2$ for the sum
of quark and antiquark polarized parton densities instead of the
valence sea quark densities, which in addition are more general,
\begin{eqnarray}
\nonumber
x(\Delta u+\Delta \bar{u})(x,Q^2_0)&=&A_{u+\bar{u}}x^
{\alpha_{u+\bar{u}}}
(1-x)^{\beta_{u+\bar{u}}}
(1+\epsilon_{u+\bar{u}}{\sqrt{x}}+
\gamma_{u+\bar{u}}x),\\[2mm]
\nonumber
x(\Delta d+\Delta \bar{d})(x,Q^2_0)&=&A_{d+\bar{d}}x^
{\alpha_{d+\bar{d}}}
(1-x)^{\beta_{d+\bar{d}}}(1+\gamma_{d+\bar{d}}x),\\[2mm]
\nonumber
x(\Delta s+\Delta \bar{s})(x,Q^2_0)&=&A_{s+\bar{s}}x^
{\alpha_{s+\bar{s}}}
(1-x)^{\beta_{s+\bar{s}}}(1+\gamma_{s+\bar{s}}x),\\[2mm]
x\Delta G(x,Q^2_0)&=&A_{G}x^{\alpha_{G}}
(1-x)^{\beta_G}(1+\gamma_{G}x),
\label{input_PDFs}
\end{eqnarray}
and do {\it not} use any assumptions about the light sea quark
densities $\Delta \bar{u}$ and  $\Delta \bar{d}$.

As usual, the set of free parameters in (\ref{input_PDFs}) is
reduced by the well-known sum rules
\begin{equation}
a_3=g_{A}=\rm {F+D}=1.269~\pm~0.003,~~~\mbox{\cite{PDG}}
\label{ga}
\end{equation}
\begin{equation}
a_8=3\rm {F-D}=0.585~\pm~0.025,~~~\mbox{\cite{AAC00}} \label{3FD}
\end{equation}
where $a_3$ and $a_8$ are nonsinglet combinations of the first
moments of the polarized parton densities corresponding to $3^{\rm
rd}$ and $8^{\rm th}$ components of the axial vector Cabibbo
current
\begin{eqnarray}
a_3&=&(\Delta u+\Delta\bar{u})(Q^2) - (\Delta
d+\Delta\bar{d})(Q^2),\\[2mm]
a_8&=&(\Delta u +\Delta\bar{u})(Q^2) + (\Delta d +
\Delta\bar{d})(Q^2)-2(\Delta s+\Delta\bar{s})(Q^2).
\end{eqnarray}

The sum rule (\ref{ga}) reflects isospin SU(2) symmetry, whereas
(\ref{3FD}) is a consequence of the $SU(3)_f$ flavor symmetry
treatment of the hyperon $\beta$ decays. So, using the constraints
(\ref{ga}) and (\ref{3FD}) the parameters $A_{u+\bar{u}}$ and
$A_{d+\bar{d}}$ in (\ref{input_PDFs}) can be determined as
functions of the other parameters connected with $(\Delta u+\Delta
\bar{u}),~(\Delta d+\Delta \bar{d})$ and $(\Delta s+\Delta
\bar{s})$.

The large $x$ behavior of the polarized PDFs is mainly determined
from the positivity constraints \cite{LSS10}, where for the
unpolarized NLO PDFs the MRST'02 set of parton densities
\cite{MRST02} has been used. In order to guarantee the positivity
condition for the polarized strange quarks and gluons we assume
the following relation for the parameters $\beta_i$ which control
their large $x$ behavior:
\begin{equation}
\beta_{s+\bar{s}}=\beta_G=\beta_{sea(MRST02)}=7.276.
\label{beta_i}
\end{equation}

The rest of the parameters $\{A_i, \alpha_i, \beta_i, \epsilon_i,
\gamma_i\}$, as well as the unknown higher twist corrections
$h^N(x)/Q^2$ to the spin structure functions $g_1^N(x,Q^2),~(N=p,
n)$ have been determined from the best fit to the DIS data. Note
that the $\sqrt{x}$ term has been used only in the parametrization
for the $(\Delta u+\Delta\bar{u})$ density, because the parameters
$\epsilon_{i}$ in front of it for the other polarized densities
cannot be determined from the fit, and do not help to improve it.
Note also that the higher twist effects are nonperturbative ones
and cannot be calculated in a model-independent way. That is why
we prefer to extract them directly from the experimental data (for
more details, see our paper \cite{HT}).

The numerical results of our NLO QCD fit to the present world data
set on polarized inclusive DIS are presented in Tables I, II and III.
\vskip 0.6cm
\begin{center}
\begin{tabular}{cl}
&{\bf TABLE I.} Data used in our NLO QCD analysis, \\ &the
individual $\chi^2$ for each set and the total $\chi^2$ of the
fit.
\end{tabular}
\vskip 0.3 cm
\begin{tabular}{|c|c|c|c|c|c|c|} \hline
~~~~~ Experiment~~~~~ &~~~~~Process~~~~&$~~N_{data}~~~$&
$~~~~\chi^2$~~~~ \\ \hline
 EMC \cite{EMC}  &    DIS(p)   &  ~10 & ~4.2 \\
 SMC \cite{SMC}  &    DIS(p)   &  ~12 & ~4.8  \\
 SMC \cite{SMC}  &    DIS(d)   &  ~12 & 17.8   \\
 COMPASS \cite{COMPASSp} & DIS(p) &  ~15 & 11.1 \\
 COMPASS \cite{COMPASSd} & DIS(d) &  ~15 & ~9.2 \\
 SLAC/E142 \cite{SLAC142}& DIS(n) &  ~~8 & ~6.7 \\
 SLAC/E143 \cite{SLAC143}& DIS(p) &  ~28 & 15.6 \\
 SLAC/E143 \cite{SLAC143}& DIS(d) &  ~28 & 39.7 \\
 SLAC/E154 \cite{SLAC154}& DIS(n) &  ~11 & ~2.0 \\
 SLAC/E155 \cite{SLAC155p}& DIS(p) & ~24 & 24.9 \\
 SLAC/E155 \cite{SLAC155d}& DIS(d) & ~24 & 16.6 \\
 HERMES \cite{HERMES} & DIS(p)  & ~~9 & ~5.1 \\
 HERMES \cite{HERMES} & DIS(d)  & ~~9 & ~5.9 \\
 JLab-Hall A \cite{JLabn}& DIS(n)& ~~3& ~0.2 \\
 CLAS'06 \cite{CLAS06} &  DIS(p) & 151 &   122.3 \\
 CLAS'06 \cite{CLAS06} &  DIS(d) & 482 &   430.0 \\
 CLAS'14 \cite{CLAS14} &  DIS(p) & ~32 &   ~17.6 \\
 CLAS'14 \cite{CLAS14} &  DIS(d) & ~29 &   ~~6.8 \\
\hline
{\bf \ TOTAL}: &        &   902 &   740.6  \\
\hline
\end{tabular}
\end{center}
\vskip 0.5cm

In Table I the data sets used in our analysis are listed and the
corresponding values of $\chi^2$ obtained from the best fit to the
data are presented. As seen from Table I, a good description of
the data is achieved: $\chi^2/{DOF}$=0.842 for 902 experimental
points using 23 free parameters (13 for the PDFs and 10 for the
higher twist corrections). The new proton and deuteron CLAS data
are well consistent with the previous world data set and very well
fitted:$\chi^2_{Nrp}=0.55$ and 0.23 per point for the proton and
deuteron data, respectively.

The values of the parameters attached to the input polarized PDFs
obtained from the best fit to the data are presented in Table II.
The errors correspond to $\Delta \chi^2=1$. Note also that only
the experimental errors (statistical and systematic) are taken
into account in their calculation. As seen from Table II, the
parameters connected with the polarized strange quark density are
well determined. Taking into account the value of the parameter
$\gamma_{s+\bar{s}}$ one sees that the  strange quark density is
\emph{negative} for small values of $x$ and changes sign in the
region $0.3<x< 0.4$ (the precise point depending on the value of
$Q^2$). Beyond this cross-over point it is exceedingly small,
compatible with zero (see Fig. 1).
\begin{figure}[h]
\centerline{ \epsfxsize=5.4in\epsfbox{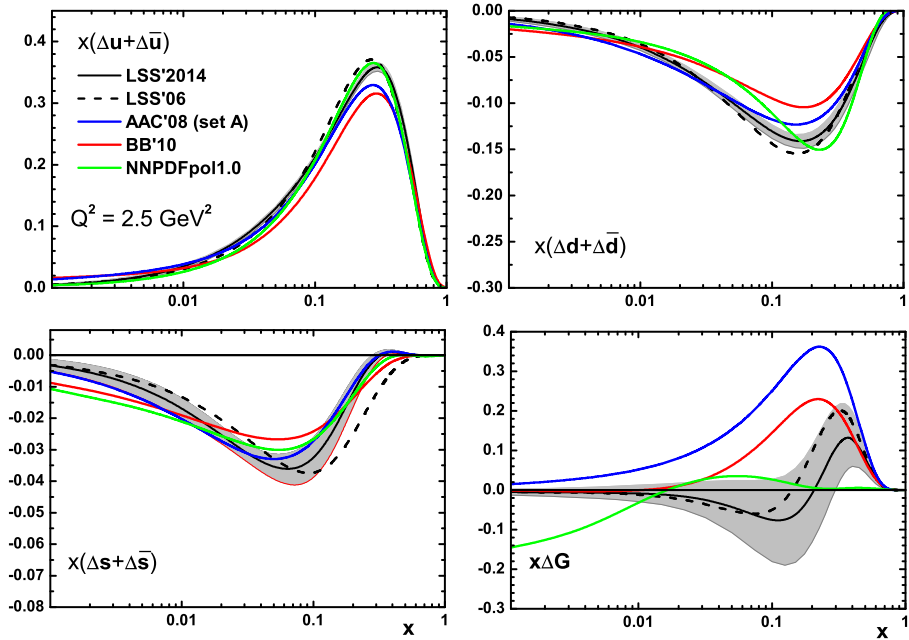} }
 \caption{Our NLO polarized PDFs compared to those of LSS'06, AAC'08,
BB'10 and NNPDFpol1.0. }
\end{figure}

The extracted polarized NLO PDFs are plotted in Fig. 1 for
$Q^2=2.5~GeV^2$ and compared to those obtained in our previous
analysis \cite{LSS07}. In Fig. 1 the AAC'08(set A), BB'10 and
NNPDFpol1.0 polarized PDFs obtained from NLO QCD analyses of the
inclusive DIS data alone (respectively the second, third and
fourth Refs. in \cite{groups}) are presented too. As seen from
Fig. 1, our new polarized parton densities (LSS'14 PDFs) are well
consistent with our LSS'06 PDFs (dashed curves). The extracted
strange quark density remains significantly negative even though
the parametrization allowed a sign change as a function of $x$
\cite{footnote}.
\begin{center}
\begin{tabular}{cl}
&{\bf TABLE II.} The parameters of the NLO input
polarized PDFs at $Q^2=1~GeV^2$ \\
&obtained from the best fit to the data. The errors shown are total
(statistical and \\&systematic).
The parameters marked by (*) are fixed.
\end{tabular}
\vskip 0.3 cm
\begin{tabular}{|c|c|c|c|c|c|c|} \hline
  Flavor &  A  &  $\alpha$ &  $\beta$ & $\epsilon$ & $\gamma$  \\ \hline
 $u+\bar{u}$& ~6.004$^*$ & 1.147~$\pm$~0.160 & 3.604~$\pm$~0.160 &
-2.389~$\pm$~0.443 & 4.207~$\pm$~0.982~ \\
 $d+\bar{d}$& -0.792$^*$ & 0.690~$\pm$~0.116 & 3.696~$\pm$~0.684 &
0 & 1.760~$\pm$~2.781  \\
$s+\bar{s}$ & -0.634~$\pm$~0.366 & 0.802~$\pm$~0.167 & 7.267$^*$ &
0 & -2.500~$\pm$~0.162 \\
G & -172.3~$\pm$~133.9 & 2.650~$\pm$~0.526 & 7.267$^*$ & 0 &
-3.659~$\pm$~1.018 \\
\hline
\end{tabular}
\end{center}
\vskip 0.5cm

We have found that the present polarized inclusive DIS data still
cannot rule out the solution with a positive gluon polarization.
The values of $\chi^2/DOF$ corresponding to the fits with
sign-changing and positive $x\Delta G(x,Q^2)$ are practically the
same: $\chi^2/DOF({\rm node}~x\Delta G )=0.842$ and
$\chi^2/DOF(x\Delta G >0)=0.845$, and the data cannot distinguish
between these two solutions (see Fig. 2 (left)). The corresponding
strange sea quark densities are shown in Fig. 2 (right). As seen,
the strange sea quark densities obtained in the fits with
sign-changing or positive gluons are almost identical. The
corresponding $\Delta u + \Delta \bar{u}$ and $\Delta d + \Delta
\bar{d}$ parton densities are not presented because they cannot be
distinguished from those corresponding to the changing in sign
gluon density.
\begin{figure}[bht]
\centerline{ \epsfxsize=2.5
in\epsfbox{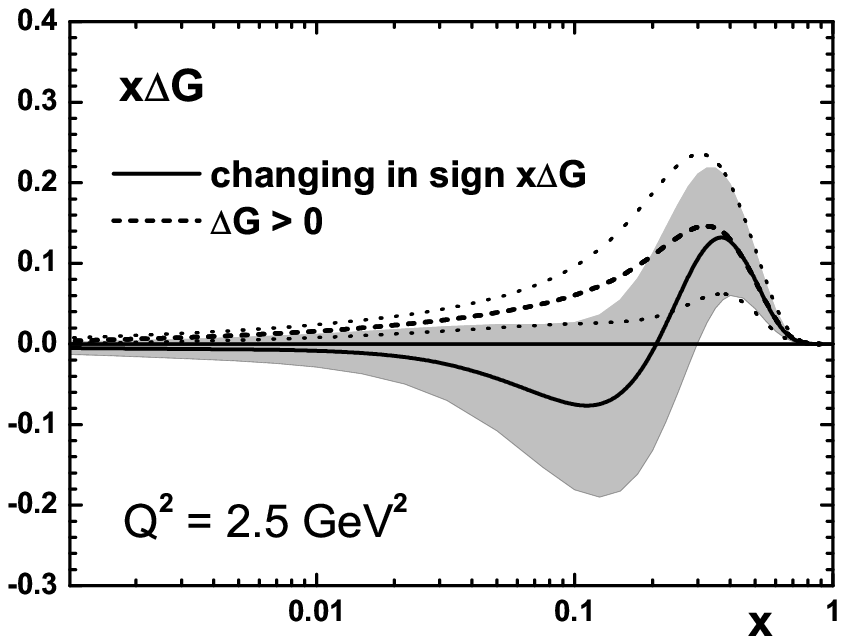}
\epsfxsize=2.5in\epsfbox{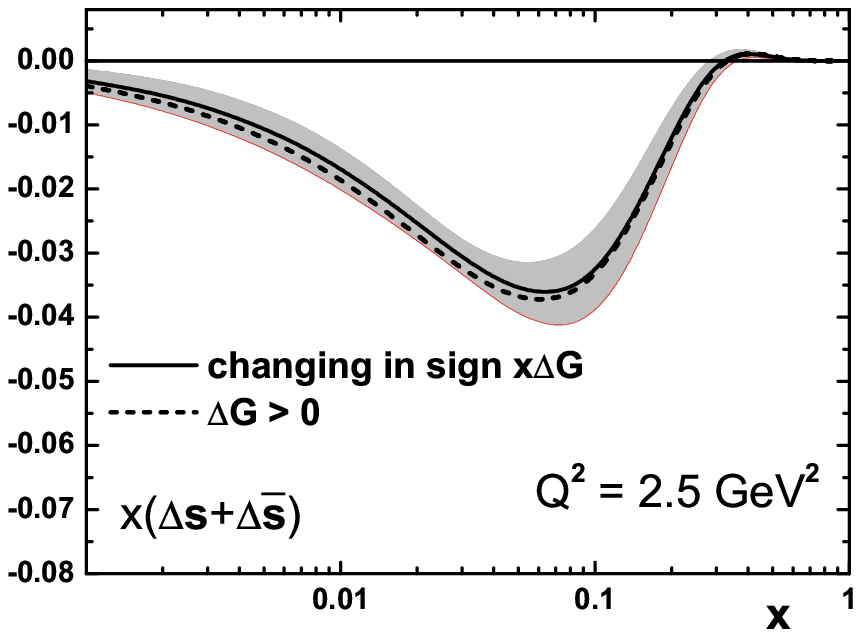} }
 \caption{Comparison between positive and sign-changing gluon densities.
The dotted curves mark the error band for positive gluons (left).
The corresponding strange quark densities are shown on the right.}
\end{figure}

In Fig. 3 our positive gluon density is compared to that obtained in
our previous analysis \cite{LSS07} when the recent CLAS data were not
available. As seen, the two gluon densities are in  good agreement.
In Fig. 3 the gluon densities obtained by AAC and BB groups are also plotted.

As was mentioned above, we take into account the higher twist
corrections to the spin structure functions in our fits to DIS
data. The values of the HT corrections $h^p(x_i)$ and $h^n(x_i)$
for the proton and neutron targets extracted from the data in this
analysis are presented in Table III. For the deuteron target the
relation $h^d(x_i)=0.925[h^p(x_i)+h^n(x_i)]/2$ have been used,
where 0.925 is the value of the polarization factor $D$.
 \begin{figure}[bht]
\centerline{ \epsfxsize=2.8in\epsfbox{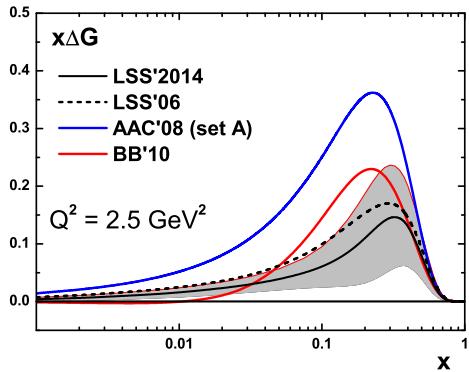} }
 \caption{Our positive solution for $x\Delta G$ compared to LSS'06,
AAC'08 and BB'10 polarized gluon densities.}
\end{figure}
\vskip 0.6cm
\begin{center}
\begin{tabular}{cl}
&{\bf TABLE III.} The values of higher twist corrections\\
&extracted from the data in a model-independent way. \\
&$<x_i>$ are the mean values of the $x_i$ bins.
\end{tabular}
\vskip 0.3 cm
\begin{tabular}{|c|c|c|c|c|c|c|} \hline
~~~~ $<x_i>~~~~$ &~~~~~ $h^p(x_i)~[GeV^2]~~~~$ &~~~~~~ $<x_i>$&~~~
$h^n(x_i)~[GeV^2]~~~$\\ \hline
 0.028 &~~~ -0.026~$\pm$~0.042 &~~~~~ 0.028 &~~~ 0.162~$\pm$~0.056 \\
 0.100 &~~~ -0.071~$\pm$~0.018 &~~~~~ 0.100 &~~~ 0.115~$\pm$~0.043 \\
 0.200 &~~~ -0.045~$\pm$~0.012 &~~~~~ 0.200 &~~~ 0.020~$\pm$~0.021 \\
 0.350 &~~~ -0.030~$\pm$~0.009 &~~~~~ 0.325 &~~~ 0.029~$\pm$~0.016 \\
 0.600 &~~~ -0.011~$\pm$~0.012 &~~~~~ 0.500 &~~~ 0.014~$\pm$~0.014 \\
\hline
\end{tabular}
\end{center}

\vskip 0.6cm

\section{Conclusion}

We have stressed that, in principle, the inclusive DIS data
uniquely determine the polarized strange  quark density. Our new
analysis of the inclusive world data, including for the first time
the extremely accurate JLab CLAS data  on the proton and deuteron
spin structure functions and the recently published COMPASS proton
data, despite allowing in the parametrization, for a possible sign
change, has confirmed the previous claim, namely, that the
inclusive data yield significantly negative values for the
polarized strange quark density. The fundamental difference
between the SIDIS and DIS analysis is the necessity in SIDIS to
use information on the fragmentation functions, which are largely
determined from multiplicity measurements. In an earlier study
\cite{LSS11} we showed that the polarized strange quark density
extracted  from SIDIS data was extremely sensitive to the input
fragmentation functions. Thus we believe that the present
disagreement between the SIDIS and DIS strange quark polarizations
very likely results from a lack of correctness of the
fragmentation functions utilized and that the results from the
inclusive analysis are correct.

\begin{center}
{\bf Acnowledgments}
\end{center}

One of us (D. S. ) is grateful to M. Hirai and S. Kumano for
providing us with their AAC'08 PDFs, as well as for the useful
discussion. We thank also J. Rojo and E. Nocera for providing us
with the NNPDFs. This research was supported by the JINR-Bulgaria
Collaborative Grant, and by the Russian Foundation for Basic
Research Grants No. 12-02-00613, No. 13-02-01005, and No.
14-01-00647.


\begin{thebibliography}{99}
\bibitem{note}
Note that the AAC group (see 2nd  Ref. in \cite{groups} and
references therein) has used a very peculiar parametrization for
the strange quark density in their pure inclusive DIS analyses
which, in principle permits the change of the sign, but one of the
parameters responsible for the sign change is fixed by hand.
\bibitem{footnote}
Note that this property of the polarized strange quark density
holds also if for the nonsinglet axial charge $a_8$ in Eq.
(\ref{3FD}), instead of its SU(3) symmetric value 0.585, the value
0.46 is used. This value corresponds to the maximal reduction of
$a_8$ presented in the literature, and is the value predicted in
one of the models on SU(3) breaking effects for $a_8$
\cite{red_a8}. The key point is that the shape of the strange
quark density is the same, however, its magnitude is approximately
halved. Details of the analysis on the sensitivity of polarized
parton densities to variation of the value of the axial charge
$a_8$ from its SU(3) symmetric value will be presented in a
forthcoming paper.
\end{thebibliography}
\end{document}